# Correlating Local Lattice Distortion with Dislocation Pinning in Refractory High-Entropy Alloys

Zhiling Luo[1], Wang Gao[1]*, Qing Jiang[1]*

[1] Key Laboratory of Automobile Materials, Ministry of Education, Department of Materials Science and Engineering, Jilin University, Changchun, 130022, China.

*Corresponding author(s). E-mail(s): wgao@jlu.edu.cn; jiangq@jlu.edu.cn.

**Abstract**

Local lattice distortion (LLD) of refractory high-entropy alloys (RHEAs) plays an essential role in mechanical properties and phase stability. However, the random distribution of multi-principal constituents of RHEAs inhibits the comprehension of LLD, although LLD is suggested to couple with chemical short-range-order (SRO). Herein, an analytical model is built to determine the site-to-site LLD of RHEAs by coupling the local lattice sites, the local size ordering in their environments and the global constituent information. By elucidating the size coupling between components, the model demonstrates that LLD exhibits a mechanism similar to the relaxation of metal surfaces. Moreover, it is found that LLD, rather than chemical SRO, serves as the origin of solid-solution strengthening and as a measure of the phase transformation in RHEAs. The scheme provides a comprehensive physical picture and offers a quantitative measurement of LLD at macro and micro scales, laying a foundation for the design of RHEAs.

## 1. Introduction

High entropy alloys (HEAs), as the next-generation superior structural material, exhibit outstanding mechanical strength [1–3], structural stability [4], catalytic activity [5–7], corrosion resistance [8,9], oxidation resistance [10], fatigue resistance and creep resistance [11,12]. These excellent properties, particularly yield strength [13–16], ductility [17] and radiation resistance [18–20], are suggested to depend on the local lattice distortions (LLD) of HEAs especially for refractory HEAs (RHEAs). However, the role of LLD in the mechanical properties of RHEAs is still controversial, since the chemical short-range-order (SRO) is also considered essential for the strengthening mechanism of RHEAs [21–23]. In particular, the site-to-site LLD is likely coupled with the chemical SRO of RHEAs [17], therefore it is essential to reveal the physical picture of site-resolved LLD as well as the differences and connections between LLD and chemical SRO. For any RHEAs, the random arrangement of multiple elements in lattice sites makes the LLD of a given element different from one site to another, which prohibits to study the determinants of LLD and their coupling rules [24,25]. Generally, measuring atomic-scale LLD is challenging for experiments [26–28] and computationally expensive for Density Functional Theory (DFT) calculations when scanning lattice sites at high throughput [29–32]. Hence, it is of vital necessity to provide a model for the site-to-site prediction of atomic-scale LLD, for further studying the structure-property relationship between LLD and macroscopic properties of RHEAs.

Many attempts based on the semi-empirical basis have been made to investigate the determinants of lattice distortion (LD) [33–35], such as the hard sphere model [36,37] and the soft sphere model [38,39]. Generally, these models only provide the properties of average lattice distortion (ALD) for a given RHEAs, obscuring the quantitative relationship between the local site information and the corresponding LLD for a given site. Undoubtedly, LLD uncovers the origin of LD at the most basic level and automatically derives the ALD, whereas its quantification scheme and physical picture are still great obstacles.

Herein, an analytical model is built to predict the complex site-to-site LLD of RHEAs by combining the site-resolved information (the neighbor number ratio of central atoms and the central atom radii) with the macro-scale constituent information (the standard deviation of constituent radii and the constituent number). LLD is determined by the size coupling effect between constitutes in RHEAs and experiences the similar mechanism as that of metal surface relaxation. Moreover, this scheme identifies the determining contribution of LLD to the dislocation pinning points, phase transition and solid-solution strengthening in RHEAs, which provides a novel physical picture for the structural and mechanical properties of RHEAs and facilitates the further design of RHEAs.

## 2. Methods

**Calculation of LLD in RHEAs.** To measure the LLD of RHEAs, the Wigner-Seitz (WS) radius was calculated for each atom, $R_{WS} = \left(\frac{3}{4\pi} V_{WS}\right)^{\frac{1}{3}}$, where $V_{WS}$ is the volume of WS cell [40,41]. Six quaternary and six quinary RHEAs were adopted, consisting of V, Mo, W, Ti, Nb, Ta, Hf and Zr, to systematically study the LLD of RHEAs. Using the Special Quasi-random Structure (SQS) method [42,43], 4 x 4 x 4 body-centered cubic (BCC) supercells were built for quaternary RHEAs and 5 x 5 x 5 BCC supercells were built for quinary RHEAs. For each RHEA, at least three SQS realizations were checked. These structures were optimized with the Vienna Ab initio Simulation Package (VASP) [44] by using the projector augmented wave method (PAW) [45] and Perdew-Burke-Ernzerhof (PBE) [46] functional. A plane-wave cutoff energy of 520 eV was used, and the conjugate gradient algorithm was utilized with a convergence threshold of 0.02 eV/Å for the Hellmann-Feynman force on each atom. The Monkhorst-Pack k-point sampling mesh density was 3 x 3 x 3 [47] and the step size was less than 0.028 Å$^{-1}$ for geometry optimization on all considered systems. In addition, 8 x 8 x 8 BCC supercells were built for large-size MoWNbTa RHEAs, which were optimized by Moment Tensor Potential (MTP) [48,49] using Large-scale Atomic/Molecular Massively Parallel Simulator (LAMMPS) code [50] with MAterials Machine Learning (MAML) package [51]. After a rigorous test, the maximum error between the $R_{WS}$ optimized by MTP and DFT is 0.002 Å, which is one order of magnitude smaller than the fluctuation range of 0.033 Å of $R_{WS}$ in MoWNbTa optimized by DFT (Fig.S1 in Supplementary Materials (SI)). Therefore, MTP has sufficient accuracy for the optimization of large-size supercells.

**Molecular dynamic simulation of edge dislocation motion.** To study the local pinning effect of LLD on edge dislocation slide in RHEAs, the Molecular Dynamics (MD) simulations were performed using LAMMPS code [50]. The edge dislocation glide in MoWNbTa and MoTiNbTa were computed via MTP potential [48,52]. The simulation cell was illustrated in Fig. S2. The X- and Y-directions were set as periodic and Z-direction was set as shrink-wrapped nonperiodic. The dislocation line was inserted in the simulation cell similar to the previous MD studies of edge dislocations in MoWNbTa RHEAs [48] and the Burgers vector was 1/2<111>. The structures were optimized by MTP with a force tolerance of $10^{-9}$ eV/Å [48]. Then, the initial velocity of the top two atomic layers was set to 0.1 Å/ps in the X-direction and the initial velocities were set to 0 in the Y- and Z-directions. The simulated temperatures were maintained at 1 K or 1200 K using the NVT ensemble. The time step was 0.001 ps during the simulation. The dislocation line structure were identified and displayed by *Dislocation Extraction Algorithm* method with OVITO code [41,53]. Notably, as the crystal structure of the dislocation line atoms is destructed, we refer to the methods of *Adaptive Common Neighbor Analysis* [14] and *Conventional Common Neighbor*

*Analysis* [15] to distinguish the neighboring atoms of the dislocation line atoms. The calculation details are shown in Note.S2.

## 3. Results

As a direct reflection of LLD, the $R_{WS}$ of RHEAs exhibits a significant discrete distribution. Taking MoWNbTa HEA as an example, the $R_{WS}$ of Ta and Nb are larger than the average $R_{WS}$, while the $R_{WS}$ of Mo and W are smaller than the average $R_{WS}$ (Fig.1A), corresponding to the fact that the atomic radii of Ta and Nb are larger than those of Mo and W. The LLD of a given atom in RHEAs is largest when the $R_{WS}$ of this atom exhibits the average of the radii of the constituent elements, e.g. the LLD is largest at $R_{WS}$ = 1.593 Å in MoWNbTa HEA shown in Fig. 1A. In addition, the $R_{WS}$ distribution of a given constituent is obviously discrete due to the variation of environments from one site to another. Therefore, the $R_{WS}$ of RHEAs is determined by the central atoms and their environments.

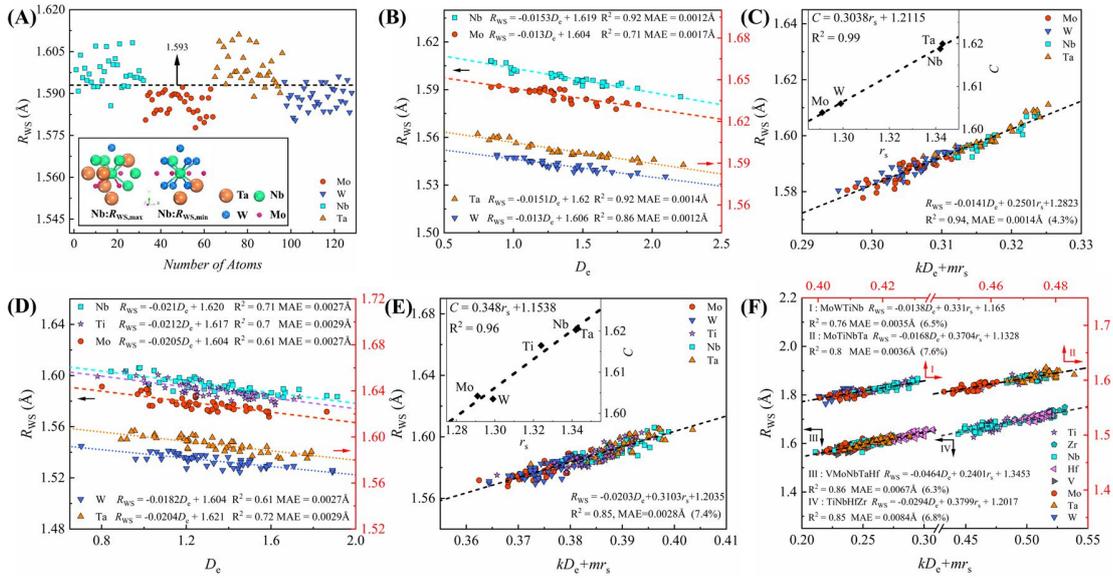

**Fig.1** The descriptors for determining the environmental effect and central-atom effect on the Wigner-Seitz radius ($R_{WS}$) of RHEAs. (A) The $R_{WS}$ distribution of MoWNbTa HEA and the average $R_{WS}$ is 1.593 Å. The inset shows the environments with the maximum and minimum radii of Nb as the central atom. The $R_{WS}$ of each element in (B) MoWNbTa and (D) MoWTiNbTa RHEAs against the environmental descriptor $D_e$. The $R_{WS}$ of (C) MoWNbTa, (E) MoWTiNbTa, (F) MoWTiNb, MoTiNbTa, VMoNbTaHf and TiNbHfZr RHEAs against the linear combination of $D_e$ and $r_s$. The insets of (C) and (E) illustrate the offset $C$ against the single-bond $r_s$.

Firstly, we focus on the environmental effect on the LLD of RHEAs. Taking the $R_{WS}$ of Nb atoms in MoWNbTa HEA as an example (Fig.1A), the large-radius neighboring atoms increase the $R_{WS}$ of Nb atoms, while the small-radius neighboring atoms reduce the $R_{WS}$ of Nb atoms. It indicates that the large- and small-radius neighboring atoms have opposite effects on the $R_{WS}$ of central atoms. Accordingly, the ratio between the number of large- and small-radius atoms in the neighbors of central atoms is used to capture the environmental effect, by building a descriptor $D_e = \left(\frac{N_{xth}^{sr}}{N_{yth}^{lr}}\right)^z$ where $N^{sr}$ and $N^{lr}$ mean the number of surrounding atoms with the small and large radius, respectively. For quaternary MoWNbTa, $N^{sr}$ is the number of Mo and W and $N^{lr}$ is the number of Nb and Ta in the neighbors of central atoms. The influence scope of small- and large-radius neighboring atoms is tested by the $x$th and $y$th nearest neighbors (nn), and the coupling strength between small- and large-radius neighboring atoms is characterized by the $z$ value. The results show that,

$$D_e = \left(\frac{N_{1nn}^{sr}}{N_{1nn+2nn}^{lr}}\right)^{\frac{1}{2}} + \left(\frac{N_{2nn}^{sr}}{N_{2nn+3nn}^{lr}}\right)^{\frac{1}{2}} \quad (1)$$

$D_e$ is linearly correlated with the $R_{WS}$ of a given element in MoWNbTa (Fig.1B), indicating that $D_e$ effectively quantifies the crucial environmental effects on the LLD of RHEAs. Notably, for a small-radius atom in RHEAs, its large-radius (small-radius) neighboring atoms lead to significant (minor) LLD. Such environmental effects indicate that $D_e$ reflects an atomic size ordering in RHEAs. Moreover, $D_e$ also reflects the fact that in the neighbors of a given central atom, the small-radius atoms have extra space for relaxation, while the large-radius atoms are pushed away and affect their next nearest neighbors. These results demonstrate the large-radius neighboring atoms have larger influence scope than the small-radius neighboring atoms, namely, the impact of the former is more non-local compared to that of the latter. In addition, the scheme shows that it is indispensable to include the effect of the large-radius atoms of the third nearest neighbors in order to accurately determine LLD, which is the first quantitative model to demonstrate that the impact range of LLD in RHEAs reaches the third nearest neighbors (about 4 Å) [30,54]. As shown in Fig.1B, $R_{WS}$ can be calculated with $D_e$, as,

$$R_{WS} = kD_e + C \quad (2)$$

The nearly parallel curves of different elements in MoWNbTa demonstrate that $k$ is approximately constant, whereas $C$ strongly depends on the central-atom effects. $C$ is found to linearly depend on the radius of constituent elements (the inset of Fig.1C), as,

$$C = mr_s + \theta \quad (3)$$

where $r_s$ is the single-bond radius obtained from the metal dimer (Table.S1) [55], which excludes the impact of multiple coordination environments on the metal atomic radius. Based on above results, the $R_{WS}$ of any central atom in MoWNbTa can be calculated by the linear combination of $D_e$ and $r_s$, as,

$$R_{WS} = kD_e + mr_s + \theta \quad (4)$$

In Fig.1C, although the $R_{WS}$ of MoWNbTa varies over a range ($V_{ra}$) of only 0.033 Å, our model effectively realizes a site-to-site determination of $R_{WS}$ by simply summing the environmental and central-atom effects. The fitting coefficient ($R^2$) is 0.94 and the mean absolute error (MAE) is 0.0014 Å that is ~4.3% of the radius range (MAE/$V_{ra}$ = 4.3%).

For a given element, its role is distinct in determining the $R_{WS}$ of different constituent RHEAs. Taking the Ti-based RHEAs as examples, Ti has a negligible effect in MoWTiNbTa (Fig.1D,E), a non-local effect in MoWTiNb and a local effect in MoTiNbTa (Fig.1F), corresponding to the fact that Ti is the middle-radius element in MoWTiNbTa, the large-radius element in MoWTiNb and the small-radius element in MoTiNbTa ($r_{Mo} < r_W < r_{Ti} < r_{Ni} < r_{Ta}$). The same behavior has also been found for Nb in VMoNbTaHf, MoWNbTa and TiNbHfZr (Fig.1F). These results demonstrate that for a given element in different constituent RHEAs, its effect on $R_{WS}$ is strongly dependent on the radius order between this element and the other constituents.

The model is general in many more RHEAs, such as quaternary and quinary RHEAs with equiatomic or unequiatomic structures (Fig.2). In all these RHEAs, the variation range of radius is very small, from 0.033 Å to 0.128 Å. Strikingly, our model exhibits consistently well linear relationships for all RHEAs, for which MAE ranges from 0.0014 Å to 0.0084 Å with MAE/$V_{ra}$ ranging from 4.3% to 7.6%. These results prove that our model is predictive to the LLD of all RHEAs, independent of constituent element and concentration.

We now study the origin of parameters, $k$ and $m$, in the Eq.(4). $k$ and $m$ are constant for a given HEAs and are distinct for different HEAs, thus $k$ and $m$ depend on the macroscopic compositional information of the HEAs. It turns out that $k$ is a linear function of the standard deviation between the $r_s$ of constituents, $r_{s-sd} = \sqrt{\sum_{i=1}^{N} c_i(r_{s,i} - \bar{r_s})^2}$, where $c_i$ is the concentration of constituent $i$ and $N$ is the number of constituents of RHEAs, and the relationship between $k$ and $r_{s-sd}$ can be simplified as $k = -\frac{3}{5}r_{s-sd}$ (Fig.3). Meanwhile, $m$ can be calculated as $1/(N-1)$, representing the average effect of a given element on the other constituents, which complies with the mean field effect.

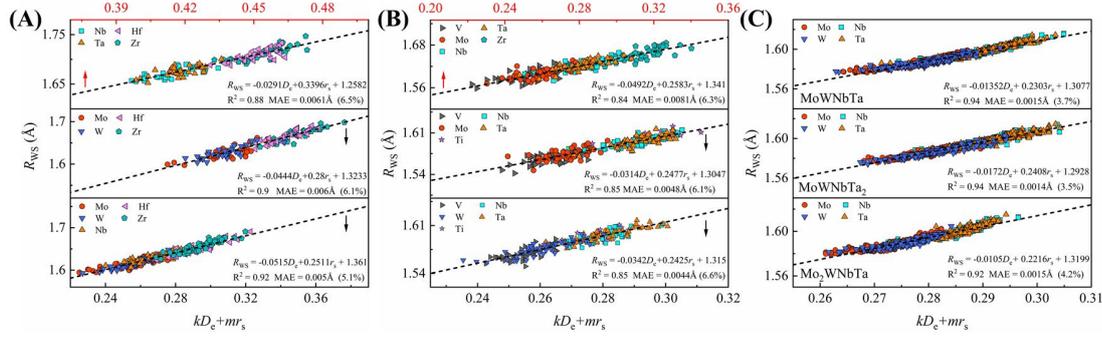

**Fig.2** The generality of the linear relationship between $R_{WS}$ and $kD_e + mr_s$. The $R_{WS}$ of (A) NbTaHfZr, MoWHfZr and MoWNbHfZr (B) VMoNbTaZr, VMoTiNbTa and VWTiNbTa RHEAs, (C) large-cell MoWNbTa, MoWNbTa$_2$ and Mo$_2$WNbTa RHEAs containing 1024 atoms, against the linear combination of $D_e$ and $r_s$.

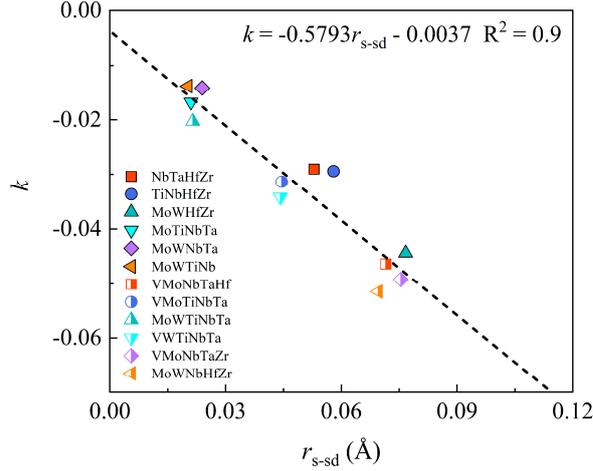

**Fig.3** The slope $k$ of the scaling relation between $R_{WS}$ and $D_e$ in different RHEAs, as a function of the standard deviation ($r_{s-sd}$) between the $r_s$ of constituents.

Based on the determinants of LLD at the micro and macro scale, a fully analytical model of $R_{WS}$ is obtained, as:

$$R_{WS} = -\frac{3}{5}r_{s-sd}D_e + \frac{r_s}{N-1} + \theta$$
$$= -\frac{3}{5}\sqrt{\sum_{i=1}^{N} c_i(r_{s,i} - \bar{r})^2}\left[\left(\frac{N_{1nn}^{sr}}{N_{1nn+2nn}^{lr}}\right)^{\frac{1}{2}} + \left(\frac{N_{2nn}^{sr}}{N_{2nn+3nn}^{lr}}\right)^{\frac{1}{2}}\right] + \frac{r_s}{N-1} + \theta \quad (5)$$

This analytical model achieves for the first time the site-to-site quantification of $R_{WS}$ in RHEAs, and predicts the $R_{WS}$ in 12 HEAs with a well precision, for which the MAE ranges from 0.0024 Å to 0.0103 Å with MAE/$V_{ra}$ ranging from 5.5% to 8.7% (Fig.S2). Notably, all involved parameters are readily accessible and can be determined based on atomic radii. The results demonstrate that the intrinsic size of elements plays a crucial role in the $R_{WS}$ of HEAs (the first two parts of the right-hand side of Eq.5), while the alloying effect has a small effect on the trend of LLD, approximated as a constant for a given HEAs, captured by the $\theta$. Additionally, the model demonstrates that $R_{WS}$ (corresponding to LLD) is determined by the coupling between the local site-resolved information (reflected by $D_e$ and $r_s$) and the global constituent information (reflected by $r_{s\text{-}sd}$ and $N$) of RHEAs. Therefore, our model not only builds a novel physical picture for LLD, but also has rich connotations to encapsulate the complexity of LLD.

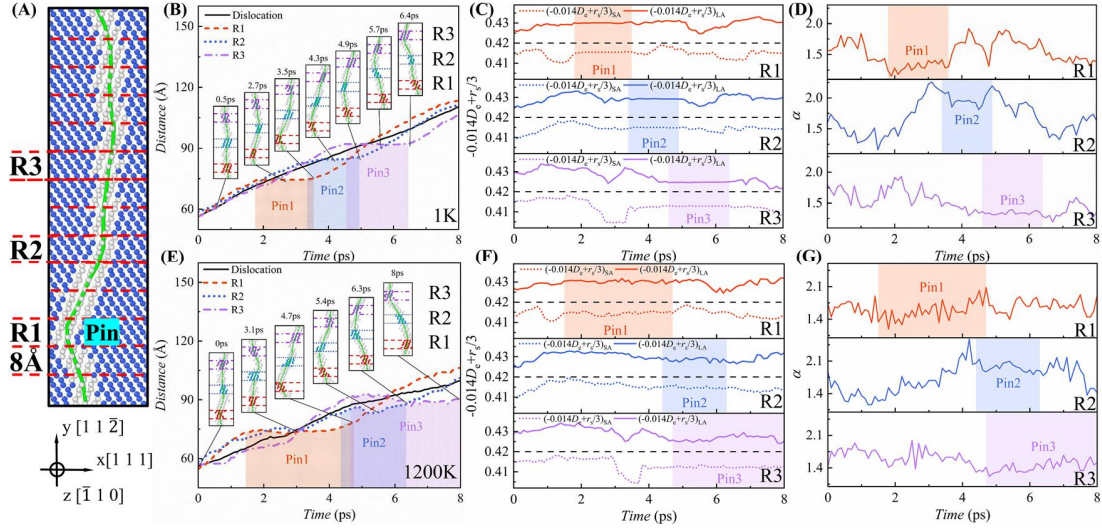

**Fig.4** The pinning points of LLD on the glide of edge dislocation in MoWNbTa RHEA. (A) The snapshot of atomic configurations with the edge dislocation (green line) gliding on {110}<111> slip system, illustrates a representative local pinning point during edge dislocation glide in MoWNbTa RHEAs. To clearly characterize the local pinning points during edge dislocation glide, the dislocation line of the simulation box is divided into 14 regions consecutively in Y-direction with the division interval of about 8 Å, and calculated the properties of edge dislocation motion as a function of simulation time for the atoms of the dislocation line of each region during edge dislocation gliding at the temperature of 1 K (B-D) and 1200 K (E-G). The gliding distance of the atoms of the dislocation line and the representative snapshots for regions 1-3 (R1-3) during dislocation gliding at 1 K in (B) and at 1200 K in (E). The black solid line is the gliding distance of the total dislocation line. Atoms are in red for R1, in blue for R2, and in purple for R3. The average LLD ($-0.014D_e + r_s/3$) in (C) and (F) and the average chemical SRO descriptor ($\alpha$) in (D) and (G) for the atoms of the dislocation line during dislocation gliding in R1-3, respectively.

## 4. Discussion

Here we turn to uncover the physical picture of LLD in RHEAs with the descriptor $D_e$, which is found to be analogous to the relaxation of metal surfaces and can be understood with the surface energy model in the tight-binding (TB) approximation [56]. It is known that for transition metals, it is sufficient to only consider the band contribution ($E_{band}$), which is proportional to the square root of the coordination number ($Z_c$), noted as $E_{band} \propto (Z_c)^{\frac{1}{2}}$. On the other hand, $E_{band}$ is related to the radius ($R$) of surface atoms of metals as: $E_{band} = C_1 \left[ e^{C_2(R/R_0 - 1)} - \frac{1}{2} \right]^{\frac{1}{2}} + C_3$ [56], where $R_0$ is the metal atomic radius in bulk. At least the second order approximation is needed to capture the relationship between radius variation and $E_{band}$. In this case, $E_{band} \propto R$, thus $R \propto (Z_c)^{\frac{1}{2}}$. In addition, for a given central atom, the large-radius neighboring atoms are pushed away and reduce its $Z_c$, whereas the small-radius neighboring atoms are brought closer and increase its $Z_c$. Accordingly, $R$ is proportional to $(N^{sr}/N^{lr})^{\frac{1}{2}}$ as $D_e$.

Benefiting from the site-to-site quantification of LLD, our model can be employed to investigate the contribution of LLD to the mechanical properties of RHEAs at the atomic scale [2,8,17], such as dislocation motion. In RHEAs such as MoWNbTa, the glide of edge dislocations plays a dominant role in the solid-solution strengthening (SSS) over a wide temperature range, especially at high temperatures [57,58]. However, the origin of pinning points during dislocation glide is a controversial but fundamental issue [48,59]. Therefore, the edge dislocation motion in MoWNbTa is studied to understand the local pinning effects in RHEAs by using MD method with machine-learning force field [48]. Fig.4 illustrates that the edge dislocation line of MoWNbTa bends during gliding, due to the local pinning points caused by intense LLD that can be determined by the descriptor $-0.014D_e+r_s/3$ of MoWNbTa. We focus on the three sequential pinning points of dislocation line gliding in region 1-3 (R1-3) at low temperature (1 K in Fig. 4B-D) and high temperature (1200 K in Fig. 4E-G). The complete pinning processes of edge dislocation glide are shown in Supplementary Movies 1 and 2. To explore the detailed effects of different-size atoms at local pinning points, $-0.014D_e+r_s/3$ for the small-radius and large-radius atoms of the dislocation line in R1-3 are calculated respectively, denoted as $(-0.014D_e+r_s/3)_{SA}$ and $(-0.014D_e+r_s/3)_{LA}$, respectively (Fig.4C and 4F). The average $-0.014D_e+r_s/3$ of edge dislocation line is the reference value (equals 0.42) to determine the largest LLD of the atoms at the edge dislocation line in MoWNbTa RHEAs, due to the destruction of the crystal structure at the edge dislocation line. First, the three sequential pinning points Pin1-3 at 1 K temperature is studied to reduce the thermal contribution (Fig.4B). Fig.4C indicates that $(-0.014D_e+r_s/3)_{SA}$ and $(-0.014D_e+r_s/3)_{LA}$ remain almost unchanged in the pinning points and are close to 0.42. Therefore, the condition for the robust local pinning effect is that small- and large-radius atoms have stable significant LLD with $(-0.014D_e+r_s/3)_{SA}$ and $(-0.014D_e+r_s/3)_{LA}$ deviating from 0.42 by < 2.4% (0.41-0.43) (Fig.4C). Then, for the same three sequential pinning points Pin1-3 at 1200 K temperature (Fig.4E), the dislocation lines are not strictly pinned due to the thermal contribution, reflected by the non-negligible fluctuating glide speed of the dislocation lines at the pinning points. Our model reveals that at high temperature, the thermal contribution makes only small-radius atoms (large-radius atoms) exhibit stable pronounced LLD with $(-0.014D_e+r_s/3)_{SA}$ $((-0.014D_e+r_s/3)_{LA})$ deviating from 0.42 by < 2.4%, and large-radius atoms (small-radius atoms) have large fluctuations (up to 0.004) (Fig.4F). Namely, only small- or large-radius atoms generate the local pinning effect. Accordingly, the glide of the dislocation line slows down significantly, but cannot form a robust pinning. These results indicate that LLD, as an intrinsic core effect of RHEAs, remains valid at high temperatures for the pinning of edge dislocations, while the thermal fluctuations weaken the formation of robust local pinning reducing the yield strength of RHEAs at high temperatures [57,60]. Moreover, our model demonstrates that LLD is determined by the atomic-size ordering of constituents in RHEAs, thus the LLD-dominated local pinning points are independent of the type and concentration of constituents. These findings can be further confirmed by the calculations of the glide of edge dislocations in MoTiNbTa, $Mo_{0.1}(WNbTa)_{0.9}$ and $Mo_{0.05}(WNbTa)_{0.95}$, where LLD indeed determines the local pinning of edge dislocation glide (see Fig.S5 and S6). These results prove that LLD is essential in controlling the local pinning points that determine the glide of edge dislocations in RHEAs.

Fig.1C shows that the sizes of small-(large-)radius atoms, Mo and W (Nb and Ta), are very close to each other, thus the average of $(-0.014D_e+r_s/3)_{SA}$ $((-0.014D_e+r_s/3)_{LA})$ of the dislocation atoms during dislocation gliding depends mainly on $D_e$ in MoWNbTa RHEA. These results prove that the size ordering is essential in controlling the local pinning points of edge dislocations glide in RHEAs. In contrast, the local pinning points cannot be captured by the chemical SRO. The widely used chemical pairwise SRO parameter are calculated, $\alpha_{ij} = 1 - \frac{P_{j,i}}{c_j}$ [23] ($P_{j,i}$ is the fraction of species j in the first nearest-neighboring shell around i, and $c_j$ is the concentration of j) of the dislocation line in R1-3 during gliding at 1 K (Fig.4D) and at 1200 K (Fig.4G). In Fig.4D, at 1 K, α cannot act as a benchmark to distinguish between pinning points and non-pinning regions. For example, α of Pin1 and Pin3 ranges from 1.15 to 1.42, while α of Pin2 ranges from 1.81 to 2.22. α is almost unchanged in Pin3, but has large fluctuations (up to 0.4) at Pin1 and Pin2 (Fig.4D). In Fig.4G, at 1200 K, the thermal contribution intensifies the oscillation of α making no significant difference between pinning points and non-pinning regions, especially for Pin1. This is due to the fact that α reflects the chemical SRO of constituents but not the size ordering, which leads to the uncertainty of α in determining the local pinning effect dominated by the size of constituent. These results are in contrast to the previous findings that the chemical SRO enhanced the movement of edge dislocations [48]. The main reason is that Ref.[48] only considered the chemical SRO of the bulk phase or the glide surface of RHEAs instead of the effect of the chemical SRO of the local pinning points in RHEAs. Additionally, the local pinning effect is also confirmed by the calculations of the glide of edge dislocations in MoWNbTa and MoTiNbTa with randomly distributed constituents (see Fig.S7). This further demonstrates the dominance of LLD in the local pinning of edge dislocation glide in RHEAs.

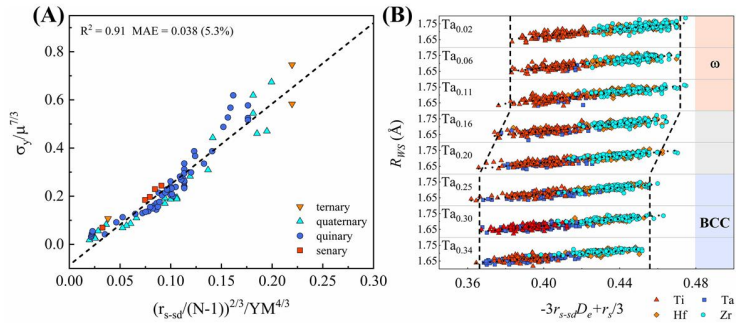

**Fig.5** LLD-determined phase transformation and yield strength of RHEAs. (A) The scaling relation between $\sigma_y/\mu^{7/3}$ and $\left(r_{s-sd}/(N-1)\right)^{2/3}/YM^{4/3}$. (B) The $R_{WS}$ of Ti$_{(1-x)/3}$Ta$_x$Hf$_{(1-x)/3}$Zr$_{(1-x)/3}$ RHEAs with x ranging from 2% to 34%, as a function of the analytical descriptor $(-3r_{s-sd}D_e/5 + r_s/3)$.

The model also provides an important strategy for the design of high-yield strength RHEAs, by modulating the macroscopic constitution of RHEAs. The atomic-scale LLD determines the local pinning points during edge dislocation glide in RHEAs, which favors the strong SSS effect corresponding to high-yield strength ($\sigma_y$) at the macro scale [15,34,35]. According to the theory of SSS for random alloys, the athermal $\sigma_y$ of a BCC solid-solution alloy can be calculated as: $\sigma_y = 3.067 A_\sigma \alpha^{-\frac{1}{3}} \mu \left(\frac{1+v}{1-v}\right)^{\frac{4}{3}} \left[\frac{\sum_i c_i \Delta V_i^2}{b^6}\right]^{\frac{2}{3}}$ [57,61], where $\alpha = \frac{1}{12}$ is a fixed line tension coefficient, v is Poisson's ratio, μ is the isotropic shear modulus and $A_\sigma$ is a constant parameter for BCC HEAs. By combining a database of experimental σ of as-cast and annealed RHEAs at room temperature (The database and reference are shown in SI) [62,63], an equation to predict the room-temperature $\sigma_y$ of RHEAs is proposed:

$$\sigma_y \approx a\mu \left(\frac{\mu}{YM}\right)^{\frac{4}{3}} \left(\frac{r_{s-sd}}{N-1}\right)^{\frac{2}{3}} \quad (6)$$

where YM is the calculated Young modulus calculated using the rule of mixtures, see the calculation detail in SI [63]. $\left(\frac{\mu}{YM}\right)^{\frac{4}{3}}$ reflect the elastic anisotropy $\left(\frac{1+v}{1-v}\right)^{\frac{4}{3}}$ calculated by v. The DFT calculated volume mismatch parameter $\left(\frac{\sum_i c_i \Delta V_i^2}{b^6}\right)^{\frac{2}{3}}$ is efficiently determined by our analytical model $\left(\frac{r_{S-sd}}{N-1}\right)^{\frac{2}{3}}$ [64]. In contrast to the atomic-scale LLD, which is determined by both the central atom and its local environment, the ALD of RHEAs at the macroscopic scale depends mainly on the number of constituents and the central atom radius. Fig.5A shows that our scheme Eq.(6) can predict the $\sigma_y$ of RHEAs, and provides an essential basis for engineering materials design. These results demonstrate that the yield strength of RHEAs is determined by ALD, which arises mainly from the central-atom effect and the mean field effect between the constituents. Therefore, our model reveals that the local dislocation pinning caused by severe LLD determines the microscopic origin of the SSS effect, while the standard deviation of constituent radii and the number of constituents determine the macroscopic behavior of the SSS effect, providing a comprehensive understanding of the critical contribution of LD to the yield strength of RHEAs.

The model is helpful to determine the phase transformation of RHEAs. Taking TiTaHfZr RHEA as an example, its phase structure transforms from ω phase to BCC phase as the concentration of Ta ($C_{Ta}$) increasing from 2 to 34% [65], which is well captured by our model. When $C_{Ta}$ increases from 2 to 11%, our descriptor $-3r_{s-sd}D_e/5+r_s/3$ of TiTaHfZr is at the range of 0.382-0.472 corresponding to ω phase, and when $C_{Ta}$ increases from 25% to 34%, $-3r_{s-sd}D_e/5+r_s/3$ is at the range of 0.366-0.456 corresponding to BCC phase (Fig.5B). However, as $C_{Ta}$ increases from 11% to 25%, the range of $-3r_{s-sd}D_e/5+r_s/3$ shifts from 0.382-0.472 to 0.366-0.456, revealing a clear transition of LLD due to the redistribution of the constituent atoms in TiTaHfZr. Therefore, the transition of LLD in TiTaHfZr, captured by $-3r_{s-sd}D_e/5+r_s/3$, determines the phase transformation from the ω phase to the BCC phase. Compared to the previous models [65,66], our model reduces extensive DFT calculations when determining this phase transformation and is convenient for application. Notably, for TiTaHfZr HEA, the variation of $-3r_{s-sd}/5$ can be ignored and $r_s/3$ remains unchanged towards the variation of $C_{Ta}$, thus LLD is dominant in the phase structure transformation reflected by the relationship between $R_{WS}$ and $D_e$. Obviously, our model provides a novel measure from the size perspective to understand the phase transformation of RHEAs, by quantifying the LLD, in contrast to the previous studies that understand the phase transformation from the energy perspective [65,66].

## 5. Conclusion

In summary, this work introduces an analytical model to site-to-site predict the discrete LLD of RHEAs, which is determined by a linear combination of the size ordering in the local environments and the central-atom radii, with the pre-factors determined by the standard deviation of constituent radii and the constituent number. Notably, LLD complies with the same rule as the relaxation of metal surfaces. The model demonstrates that LLD, rather than chemical SRO, controls the local pinning effects of edge dislocation glide in RHEAs at the atomic scale. Consequently, this work establishes a detailed scheme to unravel the underlying mechanism of SSS effects in RHEAs from micro-scale to macro-scale. Furthermore, the model provides a novel measure from the size perspective to comprehend the phase transformation of RHEAs. These findings reveal a comprehensive physical picture of LLD in RHEAs and serve as a solid basis for the rapid design of ultrahigh-performance RHEAs as all involved parameters are readily accessible.

**CRediT authorship contribution statement:**
W.G. and Q.J. conceived the original idea and designed the strategy. W.G. and Z.L. derived the models, analyzed the results, wrote the manuscript, performed the DFT calculations, drew figures and prepared the Supplementary Materials. W.G., Z.L. and Q.J. have discussed and approved the results and conclusions of this article.

**Declaration of Competing Interest**

The authors declare that they have no known competing financial interests or personal relationships that could have appeared to influence the work reported in this paper.

**Acknowledgments**

The authors would like to acknowledge the financial support from the National Natural Science Foundation of China (Nos. 22173034, 11974128, 52130101), the Opening Project of State Key Laboratory of High-Performance Ceramics and Superfine Microstructure (SKL202206SIC), the Program of Innovative Research Team (in Science and Technology) in University of Jilin Province, the Program for JLU (Jilin University) Science and Technology Innovative Research Team (No. 2017TD-09), the Fundamental Research Funds for the Central Universities, the computing resources of the High-Performance Computing Center of Jilin University, China.

**Data availability:**

Data will be made available on request.

**Appendix A. Supplementary material**

Supplementary Note.S1-2
Supplementary Fig.S1-4 and Tab.S1-2
Supplementary Movie 1
Supplementary Movie 2